\documentclass[journal]{IEEEtran}
\usepackage{graphicx}
\usepackage{amsmath}
\usepackage{booktabs} 
\usepackage{hyperref}

\begin{document}

\title{Large Language Models (LLMs) and Generative AI in Cybersecurity and Privacy: A Survey of Dual-Use Risks, AI-Generated Malware, Explainability, and Defensive Strategies}

\author{
    \IEEEauthorblockN{Kiarash Ahi\textsuperscript{1} and Saeed Valizadeh\textsuperscript{2}} \\[1.0ex]
    \IEEEauthorblockA{\textsuperscript{1}Virelya Intelligence Research Labs, San Francisco Bay Area, California \\
    \textsuperscript{2}Google, Mountain View, California \\
    \small \textsuperscript{1}\{ahi@virelya.org, kiarash.ahi@uconn.edu\}, \textsuperscript{2}\{svalizadeh@google.com, saeed.valizadeh@uconn.edu\}} \\[1.5ex]
    \rule{0.4\linewidth}{0.5pt}\\[1.5ex]
    \small This paper has been accepted as an invited paper.
}

\maketitle

\begin{abstract}
Large Language Models (LLMs) and generative AI (GenAI) systems, such as ChatGPT, Claude, Gemini, LLaMA, Copilot, Stable Diffusion by OpenAI, Anthropic, Google, Meta, Microsoft, Stability AI, respectively, are revolutionizing cybersecurity, enabling both automated defense and sophisticated attacks. These technologies power real-time threat detection, phishing defense, secure code generation, and vulnerability exploitation at unprecedented scales. LLM-generated malware alone is projected to account for 50\% of detected threats in 2025, up from just 2\% in 2021, emphasizing the need for next-generation security frameworks. This paper presents a comprehensive survey of the beneficial and malicious applications of LLMs in cybersecurity, including zero-day detection, DevSecOps, federated learning, synthetic content analysis, and explainable AI (XAI). Drawing on a review of over 70 academic papers, industry reports, and technical documents, this work synthesizes insights from real-world case studies across platforms like Google Play Protect, Microsoft Defender, Amazon Web Services (AWS), Apple’s App Store, OpenAI Plugin Stores, Hugging Face Spaces, and GitHub, alongside emerging initiatives like the SAFE Framework and AI-driven anomaly detection. We conclude with practical recommendations for responsible and transparent LLM deployment and trustworthy AI, including model watermarking, adversarial defense, and cross-industry collaboration—setting a new benchmark for rigorous, holistic cybersecurity research at the intersection of AI and threat defense—and offering a roadmap for secure, scalable LLM systems that serves as a critical reference for researchers, engineers, and security leaders navigating the complex challenges of AI-driven cybersecurity.
\end{abstract}


\section{Introduction}
The rapid evolution of artificial intelligence has placed Large Language Models (LLMs) and generative AI at the forefront of software innovation and cybersecurity transformation. Originally developed to enhance natural language understanding, LLMs such as GPT-4, PaLM, and Gemini are now widely adopted across industries to automate code generation, accelerate development workflows, and enable intelligent decision-making \cite{openai2023gpt4, google2023gemini, microsoft2023copilot}. However, this widespread adoption has created a double-edged sword: LLMs empower defenders—especially platform administrators like Google Play, Apple App Store, and other enterprise app platforms—to perform static code scanning, automate threat detection, and improve code quality in real time. Yet simultaneously, those same models are exploited by attackers to generate malware, obfuscate code, and discover vulnerabilities at scale. This duality introduces complex security and governance challenges, underscoring the urgent need for systematic analysis, responsible deployment, and robust defensive frameworks \cite{gartner2023risks}.

This paper presents a comprehensive survey of both the risks and opportunities associated with LLMs in cybersecurity. We explore their dual-use nature, recent industry and academic advances, and how both defenders and adversaries leverage these models for tasks such as code generation, malware design, zero-day detection, and DevSecOps, supported by architectural comparisons, benchmark studies, and cross-industry case examples.

To guide the reader, the paper is structured as follows: Section II lays the foundational background by reviewing existing literature on the evolution, capabilities, and early governance efforts concerning LLMs. Section III provides an in-depth analysis of LLM applicability in security, detailing their dual-use potential, specific threat vectors like AI-generated malware, the role of explainability, and emerging defensive strategies. Building upon this analysis, Section IV outlines key directions for future research essential for advancing the secure use of LLMs. The paper concludes in Section V, which summarizes the findings and proposes a governance roadmap rooted in explainability, privacy-by-design, federated learning, and compliance. By aligning defensive innovation with emerging safety standards, this paper contributes a timely framework for navigating the rising complexity of LLM-powered cybersecurity ecosystems.

\section{Background and Literature Review}

\subsection{Evolution and Capabilities of LLMs}
Large Language Models (LLMs) have evolved rapidly from their initial applications in natural language translation and generation to highly capable systems supporting complex software engineering tasks. Models such as GPT-4 and PaLM now perform code generation, refactoring, debugging, and even formal verification with increasing accuracy and fluency \cite{openai2023gpt4, narayanan2023dynamic}. These advancements are enabled by scaling transformer architectures and training on diverse programming and natural language corpora. Recent research from OpenAI and Google demonstrates how LLMs can integrate into full development pipelines, assisting with test case creation, API documentation, and dynamic bug resolution \cite{pearce2022automated, chen2023static, carlini2021leakage}.

\subsection{Security Risks and Early Governance Efforts}
The dual-use nature of LLMs has raised significant security concerns. On one hand, they can support code auditing and threat detection; on the other, they can generate obfuscated or insecure code, or be weaponized for malicious purposes. Prior work has emphasized the need for proactive safeguards, such as Brundage et al.'s recommendations on structured red teaming and audit trails, and the European Union’s Artificial Intelligence Act, which mandates risk assessments and transparency reports for high-impact models \cite{european2023ai, brundage2020trustworthy}. These frameworks aim to mitigate misuse while supporting responsible innovation.

\subsection{Ethics and Governance of Dual-Use LLMs}
Integrating LLMs into CI/CD pipelines automates crucial security tasks such as code review, threat detection, and compliance enforcement. GitLab and Azure DevOps showcase how GPT based tools can enable real-time security hardening and policy enforcement \cite{gitlab2023devsecops, azure2023sdl}. 

While the EU AI Act and the US NIST AI RMF represent significant strides, the global governance landscape for LLMs in cybersecurity remains dynamic, with other major technological regions developing their own distinct approaches. For instance, countries in Asia, such as China, Japan, South Korea, and Singapore, are actively formulating AI regulations and ethical guidelines that reflect their unique priorities. Understanding these varied international perspectives and fostering dialogue towards greater regulatory interoperability will be crucial for addressing the borderless nature of cyber threats and ensuring a globally coordinated response to the risks posed by dual-use AI \cite{european2023ai, brundage2020trustworthy, gitlab2023devsecops, azure2023sdl}.

\subsection{LLMs in DevSecOps Automation}
Empirical studies of GitHub Copilot and Microsoft Security Copilot illustrate how AI augmented developers are more efficient in detecting and resolving security flaws. These tools not only enhance productivity but also reduce the probability of vulnerabilities slipping into production code \cite{vaithilingam2022cooperative, microsoft2023copilotcase}.

\subsection{Human-AI Collaboration for Secure Development}
An important facet of human-AI collaboration in secure development involves leveraging AI models to augment human capabilities in threat detection. For instance, LLMs like VulBERTa are being fine-tuned to identify zero day vulnerabilities through pattern recognition in source code. These models outperform traditional static analyzers, significantly improving detection timelines and precision in identifying new attack vectors \cite{lisha2024benchmarking}.

\subsection{Privacy-Aware Deployment of LLMs via Federated Learning}
Privacy preserving LLM deployment strategies are increasingly relevant. Federated learning allows training across distributed devices without centralizing data, aligning with laws like GDPR. Kairouz et al. and Bonawitz et al. have demonstrated that these frameworks preserve privacy while maintaining model utility \cite{kairouz2021federated, bonawitz2019federated}.

\subsection{Explainability and Trust in AI Driven Defense}
The adoption of LLMs in automated security systems demands transparency. Explainable AI (XAI) methods like SHAP and LIME have been customized to make LLM based vulnerability classifications interpretable. These models help developers and analysts understand the rationale behind predictions, supporting auditability and compliance \cite{ribeiro2016why, samek2021explainable}.

\subsection{Adversarial Attacks and Model Vulnerabilities}
The integration of LLMs into security critical domains has exposed them to sophisticated adversarial attacks. Carlini et al. highlighted how training data could be extracted from LLMs, undermining confidentiality \cite{carlini2021leakage}. Wallace et al. demonstrated that prompt injection and adversarial fine tuning can manipulate LLM outputs, evading content filters. Recent work by Jia et al. organized a global competition revealing how LLMs can be tricked into generating offensive content and misinformation, emphasizing the need for rigorous adversarial testing frameworks \cite{jia2024global}.

\section{Analysis of LLM Applicability in Security}
As LLMs become deeply embedded in software development and cybersecurity pipelines, their dual-use potential has triggered increasing scrutiny. A growing body of research has documented how these models can unintentionally or deliberately produce insecure code, including cryptographic flaws, SQL injection vectors, and XSS vulnerabilities \cite{zhang2020policy, ghaffarinia2022static, jiang2019fake}. More alarmingly, the accessibility of LLMs has democratized the creation of deceptive content—enabling non-experts and malicious actors alike to generate phishing apps, polymorphic malware, and social engineering scripts at scale \cite{cybersecurity2025trends, amazon2023automated, ibm2023watsonx}. These developments reflect not isolated failures but systemic risks introduced by generative models when deployed without sufficient constraints. This section analyzes such risks through three lenses: (1) the emerging threat landscape shaped by misuse and amateur error, (2) industry-led defense strategies to mitigate LLM-enabled attacks, and (3) the broader governance and technical challenges that complicate safe deployment.

\subsection{Amateur Developers and Security Risks}
While LLMs empower rapid software creation, they have also unintentionally enabled a wave of insecure development by amateur coders. These models lower technical barriers to entry, allowing individuals with minimal security training to generate functional code quickly. However, this ease often comes at the cost of safety. Studies have shown that inexperienced developers frequently incorporate LLM-generated snippets directly into applications without validating correctness, context, or security implications \cite{palantir2023aip, ieee2023fairness}. As a result, common vulnerabilities such as improper authentication, insecure API usage, and unsafe cryptographic practices proliferate in production software.

This trend is particularly concerning in open-source and mobile app ecosystems, where low-friction publication processes allow insecure code to reach wide audiences. Unlike deliberate attacks, these security flaws emerge from structural gaps—lack of tooling, review, and awareness—highlighting the need for LLM-integrated guardrails that can proactively flag unsafe patterns for novice users. While amateur misuse stems from lack of expertise, the deliberate exploitation of LLMs by adversaries reveals a more calculated—and scalable—weaponization of generative AI.

\subsection{Malicious Actors Leveraging LLMs}
In contrast to accidental misuse by amateurs, malicious actors are leveraging LLMs as force multipliers for intentional cyberattacks. These adversaries use generative models to automate large-scale creation of malware, phishing payloads, ransomware variants, and code obfuscation strategies. Unlike conventional malware authors who required domain expertise, attackers can now prompt LLMs to output malicious scripts with minimal effort—dramatically accelerating development cycles. Industry reports indicate a sharp escalation in LLM-facilitated threat activity, with LLM-generated or assisted malware constituting a significant share of all new threats in 2025 \cite{nist2023rmf, wef2023standards, google2023safe}. Attackers further exploit LLMs to bypass static filters by generating code that mutates slightly on each iteration—evading signature-based detection systems. This reflects a paradigm shift in threat scalability: what was once human-limited is now AI-augmented, enabling adversaries to operate at industrial scale.

\subsection{Statistical Overview of LLM Related Malware}
The proliferation of LLM technologies has corresponded with a measurable increase in their exploitation by malicious actors. Recent cybersecurity reports reveal a sharp upward trend in malware generated or assisted by LLMs, raising concerns about automated threat scaling and democratized access to advanced attack tools. Table \ref{tab:malware_growth} presents a year-over-year breakdown of total malware cases and the proportion attributed to LLM-generated threats between 2021 and 2025 \cite{cybersecurity2024almanac}.

\begin{table}[htbp]
\caption{Growth of LLM-Generated Malware (2021–2025)}
\label{tab:malware_growth}
\centering
\begin{tabular}{cccc}
\toprule
\textbf{Year} & \textbf{Annual Malware} & \textbf{LLM-Assisted} & \textbf{LLM-Assisted} \\
 & \textbf{Detections (M)} & \textbf{Malware (\%)} & \textbf{Malware (M)} \\
\midrule
2021 & 83.3  & 2\%  & 1.666  \\
2022 & 104.5 & 5\%  & 5.23   \\
2023 & 150.0 & 15\% & 22.5   \\
2024 & 184.3 & 30\% & 55.29  \\
2025 & 221.2 & 50\% & 110.6  \\
\bottomrule
\end{tabular}
\end{table}

\begin{figure}[htbp]
    \centering
    \includegraphics[width=\linewidth, alt={Bar chart showing estimated global annual malware cases stacked with rapid growth of LLM-assisted malware rising from 2 percent to 50 percent between 2021 and 2025}]{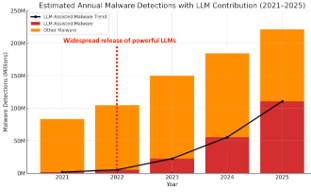}
    \caption{Estimated annual global malware detections with LLM-assisted contribution (2021–2025). Stacked bars show total malware cases, with the red portion representing LLM-assisted threats. The black line highlights the rapid growth of AI-driven malware, rising from 2\% to 50\% of all detections over the five-year period.}
    \label{fig:malware_trend}
\end{figure}

To quantify the accelerating impact of LLMs on the threat landscape, we analyzed data from Cybersecurity Ventures covering malware trends from 2021 to 2025. As shown in Figure \ref{fig:malware_trend}, total malware cases have steadily increased over this period. More notably, the share of malware attributed to LLMs has surged—from just 2\% in 2021 to a projected 50\% in 2025. This trend highlights a fundamental shift: LLMs are no longer fringe tools for experimentation, but are now actively shaping the scale, speed, and sophistication of cyber threats. The sharp growth curve underscores the need for proactive defense mechanisms that account for AI-assisted attack vectors and evolving adversarial capabilities.

\subsection{Defensive Utilization of LLMs in Mobile App Security}
In response to rising AI-powered threats, mobile platform providers are embedding LLMs directly into their security workflows. One of the most effective use cases is automated code review—where LLMs augment traditional static analyzers by identifying logic flaws, unusual API usage, or obfuscated payloads that escape signature-based detection. Google’s Gemini, for instance, plays a pivotal role in powering Play Protect, which scans millions of Android applications daily for malware, policy violations, and suspicious behaviors. By using LLMs, Play Protect has reduced both false positives and time-to-detection, allowing for proactive app ecosystem defense at unprecedented scale \cite{google2025gemini, microsoft2025copilotfaq, amazon2025whisperer, mulki2025systems}. These use cases demonstrate how LLMs can shift mobile app security from reactive filtering to intelligent pre-deployment screening, flagging issues before users ever download an app. However, as defensive applications of LLMs grow more powerful, they also inherit risks such as overfitting, bias, or exploitability—making explainability and continuous retraining essential.

\subsection{Industry Case Studies: Leveraging LLMs for Cyber Defense}
As threats fueled by LLMs escalate, leading technology companies are responding by deploying their own LLM-powered tools to reinforce digital defenses. These platforms integrate LLMs into core security operations such as code review, static analysis, compliance auditing, and threat intelligence. Each organization tailors its LLM deployment strategy to align with its security priorities, infrastructure, and customer-facing services \cite{mulki2025systems, networks2024fairness, mohindroo2024privacy, qualys2025llm, networks2025explainable}. Table \ref{tab:industry_cases} summarizes several of these leading companies and their applications of LLM technology for cyber defense.

\begin{table}[htbp]
\caption{Leading Companies Leveraging LLMs For Security}
\label{tab:industry_cases}
\centering
\begin{tabular}{lll}
\toprule
\textbf{Company} & \textbf{LLM Technology} & \textbf{Application} \\
\midrule
Google     & Gemini         & Malware Detection, Static Analysis \\
Microsoft  & GPT-4           & Security Copilot, Code Review \\
Amazon     & CodeWhisperer  & Vulnerability Detection \\
IBM        & Watsonx        & Compliance \& Risk Management \\
Palantir   & AIP            & Threat Hunting \& Behavioral Analysis \\
\bottomrule
\end{tabular}
\end{table}

These LLM-powered systems represent a shift from reactive to proactive security postures. For instance, Google's Gemini underpins Play Protect’s live threat detection engine, capable of analyzing millions of apps for suspicious behavior in real time. Microsoft's Security Copilot assists developers and analysts by flagging unsafe code patterns and generating remediation steps. Amazon’s CodeWhisperer is deeply embedded in IDEs, helping developers identify insecure code at the moment of creation. By embedding LLMs into their security stacks, these organizations are not only protecting their own platforms but also setting new industry standards for AI-augmented cybersecurity. However, the same capabilities—if left unchecked or open-sourced without safeguards—can empower adversaries, reinforcing the paper’s core thesis: LLMs are a powerful but inherently dual-use technology.

\subsection{Bias and Fairness Issues in LLM-Based Security Systems}
While LLMs offer powerful advantages in automating security analysis, they also introduce systemic risks related to bias and fairness. These models are trained on massive datasets that often reflect historical imbalances, implicit stereotypes, or geographic skew—issues that can propagate into downstream security decisions. In high-stakes environments such as app store moderation, code review, or vulnerability triage, biased outputs can result in misclassifications, disproportionately affecting certain developer communities or categories of software \cite{zafar2017fairness, barocas2019fairness}.

For example, security LLMs trained primarily on English-language or Western-centric data may struggle to accurately interpret or evaluate apps developed in other locales, leading to higher false positive rates. Similarly, bias in labeling training data (e.g., which code patterns were marked as malicious or benign) can skew the model’s risk assessments, potentially flagging harmless applications as threats or overlooking real vulnerabilities in less represented codebases. These challenges illustrate yet another edge of the sword: even defensive LLM systems can inadvertently create harm if deployed without fairness audits, dataset transparency, and debiasing techniques. As LLMs continue to be integrated into security workflows, algorithmic accountability must become a core design principle, not an afterthought.

\subsection{Scalability Concerns in LLM-Based Security Systems}
As LLMs are increasingly integrated into security pipelines, scaling these models to handle production-level traffic—especially in global platforms like app stores or CI/CD environments—presents major technical and operational challenges. Unlike isolated developer tools or research prototypes, real-world deployment demands low-latency inference, cost-effective infrastructure, and high throughput across diverse languages and architectures \cite{ahi2025gpu}.

For example, scanning millions of apps in Google Play or Apple’s App Store for policy violations, malware, or misconfigurations using LLMs requires robust resource allocation strategies, distributed model serving, and dynamic workload balancing. The complexity compounds further when real-time threat detection is needed—where every inference must complete within milliseconds, and models must be resilient to edge cases and adversarial inputs. Additionally, maintaining consistent LLM behavior across regions with different regulations (e.g., GDPR in the EU, CCPA in California) adds operational overhead. Model fine-tuning or rule enforcement may need to be localized, increasing the burden of deployment and monitoring. This raises a critical tension: LLMs are powerful, but not trivially scalable. Their integration into global security infrastructures must be carefully engineered to avoid performance bottlenecks, regional inconsistencies, and unexpected failure modes—especially as adversaries attempt to exploit system blind spots.

\subsection{Regulatory Compliance and Privacy Constraints}
The deployment of LLMs in security workflows introduces complex compliance challenges, particularly under data protection frameworks such as the General Data Protection Regulation (GDPR) and the California Consumer Privacy Act (CCPA) \cite{eu2016gdpr, ca2018ccpa}. These regulations impose strict requirements around data minimization, user consent, data residency, and the right to explanation—all of which impact how LLMs can be trained, fine-tuned, and applied to sensitive user content.

For example, static code scans or behavioral analysis performed by LLMs may inadvertently process personally identifiable information (PII) or usage metadata, triggering legal obligations. Federated learning and on-device inference offer promising paths forward, but adopting privacy-preserving techniques at scale remains technically demanding and legally ambiguous. Moreover, transparency requirements—such as explaining how a model reached a decision or why a threat was flagged—can be difficult to fulfill given the black-box nature of many large transformer models. Without rigorous documentation, organizations risk regulatory noncompliance, reputational harm, or unintended discriminatory outcomes. To truly harness LLMs for security in regulated environments, defenders must embed privacy-by-design principles into every stage of model development and deployment, while simultaneously investing in robust auditability and consent-driven architectures.

\subsection{Explainability and Trust in AI-Driven Defense}
As LLMs take on increasingly autonomous roles in cybersecurity—classifying vulnerabilities, triaging threats, or flagging anomalies—the need for explainable artificial intelligence (XAI) has become paramount. Without transparency into how these decisions are made, stakeholders may lose confidence in AI-driven defense systems, especially when they impact compliance, reputation, or user rights.

\begin{figure}[htbp]
    \centering
    \includegraphics[width=\linewidth, alt={Flowchart matrix mapping categories of explainability tools like SHAP LIME and CySecBench across security pipelines for training evaluation and compliance environments}]{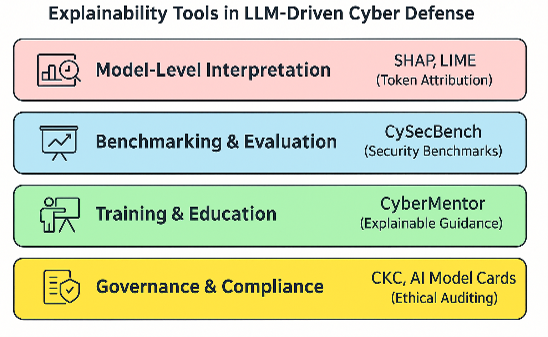}
    \caption{Categorization of explainability tools used in LLM-driven cybersecurity systems. Each pair highlights a class of explainability objective—ranging from model-level interpretation to governance—and maps it to real-world tools such as SHAP, CySecBench, and AI model cards. These tools support transparency, auditability, and trust across the security pipeline, helping address risks introduced by the opaque nature of large language models.}
    \label{fig:explainability_tools}
\end{figure}

To bridge this gap, researchers have adapted traditional XAI techniques such as SHAP and LIME to LLMs, enabling visibility into influential tokens, attention patterns, and decision pathways \cite{silva2025explainable}. These interpretations not only enhance trust but also help security analysts validate model behavior, identify edge-case failures, and fine-tune thresholds for deployment. The major categories of explainability tools and their use cases in security pipelines are summarized in Figure \ref{fig:explainability_tools}.

Recent work has also led to the creation of domain-specific benchmarks for evaluating explainability in security contexts. One such effort, CySecBench by Mishra et al. \cite{mishra2025cysecbench}, provides over 12,000 cybersecurity-focused prompts categorized by attack type, used to test how well LLMs maintain interpretability under adversarial pressure. Through prompt obfuscation and targeted jailbreaking scenarios, the benchmark has revealed varying degrees of robustness and transparency among leading models like ChatGPT, Gemini, and Claude—underscoring the urgent need for explainability tools tailored to high-risk applications.

In educational and industrial settings, tools like CyberMentor \cite{wang2025cybermentor} demonstrate how explainable AI can support cybersecurity training by offering personalized, interpretable feedback. By leveraging Retrieval-Augmented Generation (RAG) and agentic workflows, these systems teach not just what the threat is—but why it matters and how it works. More broadly, the dual-use nature of LLMs has prompted calls for ethical auditing frameworks where explainability becomes a central pillar of security governance. Strategies such as the Cyber Kill Chain (CKC) and AI model cards are being used to document vulnerabilities, decision logic, and misuse potential in a structured, auditable format. As emphasized by Barrett et al. \cite{barrett2023ethical} and Gupta et al. \cite{gupta2023framework}, explainability is not merely a UX feature—it’s essential infrastructure for regulatory compliance, misuse prevention, and long-term trust in AI-powered defense.

\subsection{Federated Learning and Privacy-Aware Deployment of LLMs}
As LLMs increasingly interact with sensitive user data—particularly in mobile, edge, and distributed environments—ensuring privacy without compromising performance has become a top priority. Federated Learning (FL) offers a promising paradigm by enabling LLM training across decentralized devices without transferring raw data to centralized servers. This approach inherently aligns with data protection regulations like GDPR and CCPA, which emphasize data locality, minimization, and user consent \cite{kairouz2021federated}.

\begin{figure}[htbp]
    \centering
    \includegraphics[width=\linewidth, alt={Block diagram comparing Centralized architecture where data goes to a single cloud-based LLM vs Federated architecture where user machines process local models and transfer encrypted parameter updates instead}]{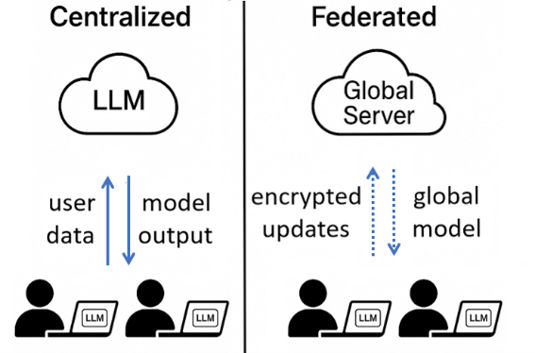}
    \caption{Architectural comparison between centralized and federated LLM deployment. In centralized systems, user data is transmitted directly to a cloud-based LLM for processing—raising privacy, security, and compliance risks. In contrast, federated learning allows users to train models locally and share only encrypted model updates with a global server, preserving data locality and enabling privacy-aware AI deployment.}
    \label{fig:federated_architecture}
\end{figure}

Kairouz et al. \cite{kairouz2021federated} provided a foundational analysis of FL's scalability and security trade-offs in privacy-critical domains, while Bonawitz et al. \cite{bonawitz2019federated} demonstrated its real-world implementation at scale within Google’s ecosystem. Their work on secure aggregation protocols—ensuring encrypted gradient updates across millions of devices—laid the groundwork for privacy-preserving AI in consumer-facing applications. By integrating LLMs with FL infrastructure, security tools can now perform real-time anomaly detection, threat classification, and on-device code analysis—without ever transmitting user data to the cloud. This architecture minimizes the risk of centralized data breaches and promotes compliance-by-design in regulated environments. Figure \ref{fig:federated_architecture} illustrates the architectural difference between centralized and federated LLM deployments, emphasizing how FL preserves data privacy by avoiding raw data transmission.

In practice, hybrid models are emerging that combine FL with on-device fine-tuning, allowing devices to benefit from shared intelligence while customizing insights for local threats. For instance, mobile app security platforms using edge-deployed LLMs can detect suspicious behaviors based on local telemetry—without exposing private logs or PII to external servers. Complementary techniques such as differential privacy, homomorphic encryption, and secure multi-party computation further harden FL pipelines, defending against inference attacks and model inversion threats. These layered approaches ensure not only privacy and accountability, but also robustness against increasingly sophisticated adversaries. In the broader security landscape, federated learning represents a critical enabler—allowing defenders to leverage the full power of LLMs while navigating the legal, ethical, and technical constraints of real-world deployment. It is a cornerstone of trustworthy AI: balancing utility, compliance, and user-centric privacy in the face of rising cyber risks.

\subsection{Detection of Zero-Day Vulnerabilities}
Zero-day vulnerabilities—unidentified, unpatched flaws that can be exploited before developers are even aware of them—pose one of the most severe threats to digital infrastructure. Traditional detection systems, which rely heavily on known attack signatures, rule-based heuristics, or static analysis, are often blind to these emerging exploits. In contrast, Large Language Models (LLMs) have shown extraordinary potential in identifying such vulnerabilities through semantic code understanding, anomaly detection, and context-aware reasoning.

In a benchmark study by Lisha et al. \cite{lisha2024zero}, a range of LLMs—including GPT-based models and fine-tuned domain-specific variants—were tested across unstructured codebases for their ability to detect zero-day vulnerabilities. The results showed that LLMs trained on vulnerability-tagged corpora and contextual embeddings significantly outperformed conventional static analyzers, particularly in uncovering logic flaws, buffer overflows, and subtle control-flow vulnerabilities in novel software. More advanced detection systems now hybridize LLMs with symbolic execution engines and graph-based models to analyze control and data dependencies. These systems can not only flag potentially vulnerable code but also hypothesize how an exploit might propagate at runtime. This enables security teams to receive both alerts and plausible exploit paths, greatly enhancing triage speed and remediation accuracy.

\begin{figure}[htbp]
    \centering
    \includegraphics[width=\linewidth, alt={Flow diagram contrasting traditional code detection pipelines based on static analysis rules with semantic code models, showing earlier lifecycle resolution within deployment environments}]{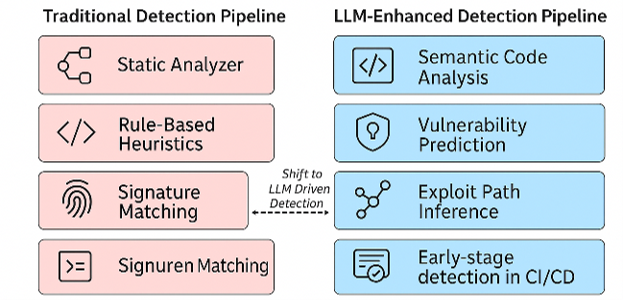}
    \caption{Comparison between traditional and LLM-enhanced zero-day vulnerability detection pipelines. Traditional approaches rely on static analyzers, rule-based heuristics, and known signatures, limiting their ability to detect unknown threats. In contrast, LLM-based systems leverage semantic code understanding, predictive modeling, and exploit path inference to detect vulnerabilities earlier and with greater accuracy.}
    \label{fig:zeroday_pipeline}
\end{figure}

In real-world applications, these techniques are being embedded directly into CI/CD pipelines. For instance, GitHub’s code scanning tools and Google’s Play Protect are experimenting with LLM-powered models that detect anomalies even in compressed or obfuscated binaries. Beyond detection, these models are also applied in fuzzing—automatically generating exploit-oriented test cases to expose weaknesses preemptively. Figure \ref{fig:zeroday_pipeline} summarizes the differences between traditional detection pipelines and LLM-based approaches, highlighting the enhanced capabilities introduced by LLMs.

Ultimately, this capability reinforces the dual-use nature of LLMs: while defenders gain new tools for anticipating and neutralizing unknown threats, attackers could also fine-tune LLMs to identify and exploit zero-day opportunities faster than ever. The same deep semantic power that enables proactive security also raises the stakes—making zero-day detection a critical battleground in AI-driven cyber defense.

\subsection{LLMs in DevSecOps Automation}
As software delivery accelerates, security must evolve to match the speed of continuous integration and deployment. DevSecOps—the integration of security directly into DevOps workflows—demands automation, precision, and scale across the entire software development lifecycle. LLMs are increasingly being leveraged to meet this need, embedding intelligence into every stage of the pipeline.

In modern DevSecOps environments, LLMs assist in:
\begin{itemize}
    \item Code scanning at every commit, flagging insecure patterns and suggesting remediations in real time.
    \item Assessing containerized builds for compliance with internal and external security policies.
    \item Analyzing dependency trees to identify vulnerable or outdated libraries before code reaches production.
\end{itemize}

Prominent platforms have already begun integrating these capabilities. GitLab’s Auto DevSecOps system employs GPT-based models for dynamic scanning and compliance-as-code enforcement. Similarly, Microsoft’s Azure DevOps, in collaboration with OpenAI, leverages LLMs for predictive vulnerability scoring, contextual remediation advice, and automated security testing. These integrations shift security from a reactive checkpoint to a proactive, continuous layer—built directly into the tooling developers already use. This minimizes friction, shortens feedback loops, and enables security-by-default at scale. At the same time, this growing reliance on LLMs in DevSecOps pipelines highlights the broader theme of this paper: the dual-use nature of AI in security. The same models that harden pipelines could be exploited if misconfigured, biased, or insufficiently governed—making LLM observability, explainability, and governance as important as their functional accuracy.

\subsection{Ethics and Governance of Dual-Use LLMs}
As LLMs continue to scale in capability, their misuse potential grows in lockstep with their utility. This presents a classic dual-use dilemma: the same model that powers security auditing, malware detection, or automated code remediation can also be harnessed to generate polymorphic malware, optimize phishing campaigns, or obfuscate malicious logic. Such high-stakes symmetry demands governance frameworks as advanced and adaptable as the technologies they aim to regulate.

Brundage et al. \cite{brundage2020trustworthy} have proposed concrete mechanisms to address these risks, including:
\begin{itemize}
    \item Structured red teaming to stress-test model behavior against adversarial use cases,
    \item Staged release strategies to control the dissemination of high-risk capabilities, and
    \item Model evaluation cards to document known limitations, safety constraints, and training data provenance.
\end{itemize}
Similarly, the strategic integration of user-experience (UX) centric human-in-the-loop (HITL) systems, drawing from principles that enhance AI-assisted productivity and decision-making, provides a critical layer of oversight for managing the operational risks of dual-use LLMs \cite{ahi2025spie}.

These ideas are now being codified in policy. The EU AI Act and the U.S. NIST AI Risk Management Framework both call for transparency in model development, auditability of training datasets, and clarity on downstream applications. These mandates aim to shift AI deployment from a reactive posture to one of accountability-by-design.

Ethical AI research further emphasizes value alignment, especially in security-critical domains. Techniques like Reinforcement Learning with Human Feedback (RLHF) are being adapted not only to optimize helpfulness, but also to enforce social norms—teaching LLMs to:
\begin{itemize}
    \item Reject harmful or manipulative queries,
    \item Disclose uncertainty in high-risk scenarios, and
    \item Explain security decisions with interpretable confidence bounds.
\end{itemize}

At the international level, coalitions such as the Global Partnership on AI (GPAI) and recent AI safety summits have introduced shared guardrails, including:
\begin{itemize}
    \item Pre-registration of frontier models,
    \item Mandatory incident reporting, and
    \item Centralized auditing repositories to detect and flag unsafe usage patterns.
\end{itemize}

\begin{figure}[htbp]
    \centering
    \includegraphics[width=\linewidth, alt={Timeline progression showing foundational AI policy milestones from 2023 up through 2025 spanning the EU AI Act NIST risk parameters and global safety summits}]{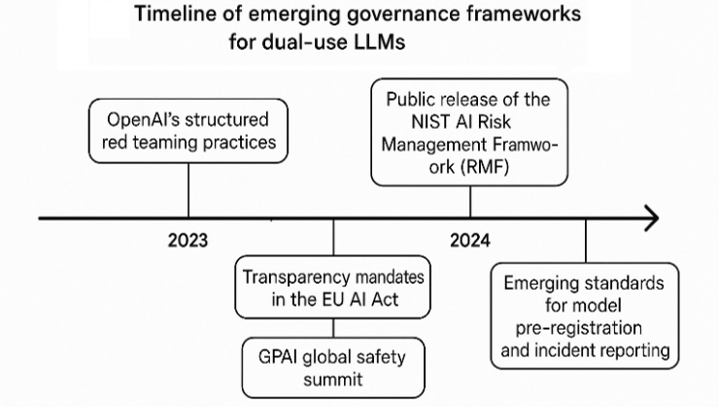}
    \caption{Timeline of emerging governance frameworks for dual-use LLMs, spanning initiatives from 2023 to 2025. Key milestones include OpenAI’s structured red teaming practices, the public release of the NIST AI Risk Management Framework (RMF), transparency mandates in the EU AI Act, the GPAI global safety summit, and emerging standards for model pre-registration and incident reporting. Together, these efforts represent a global shift toward enforceable AI safety, accountability, and dual-use risk mitigation.}
    \label{fig:governance_timeline}
\end{figure}

These governance efforts are not merely bureaucratic safeguards—they are essential infrastructure for responsibly integrating LLMs into national security, digital forensics, and trust-sensitive ecosystems. Without them, the same tools designed to protect could be weaponized—turning shields into swords. Figure \ref{fig:governance_timeline} illustrates the timeline of key governance milestones that have emerged between 2023 and 2025, highlighting a growing global effort to institutionalize safety practices around powerful LLMs.

\subsection{Illustrative Examples of LLM Exploitation and Defense}
While the adoption of LLMs by defensive platforms is notable (as discussed in Section III.E), the theoretical risks of malicious LLM use are also beginning to manifest in observable incidents. Though comprehensive public data remains scarce due to the sensitive nature of such events, emerging reports and security analyses offer early glimpses into how LLMs are being weaponized and, in some instances, how novel defenses are responding \cite{gartner2023risks, cybersecurity2025trends}. 

For example, security analysts have reported increasingly sophisticated spear-phishing emails whose linguistic complexity and contextual relevance suggest LLM assistance, bypassing conventional filters \cite{china2023interim}. Similarly, instances of novel malware variants exhibiting polymorphic behaviors potentially crafted or refined by generative models have been noted \cite{japan2022strategy}, posing new challenges for signature-based detection systems. These nascent examples underscore the practical urgency of the risks discussed and the need for continuous vigilance and innovation in defensive strategies.

\subsection{Securing Defensive LLM Systems}
As LLMs become integral components of cybersecurity infrastructure itself (e.g., in threat detection, code analysis, and incident response), their own security posture becomes paramount. Protecting these 'defender' LLMs from targeted attacks is crucial to maintain their efficacy and trustworthiness. Key considerations in safeguarding these sentinel AI systems include:
\begin{itemize}
    \item \textbf{Training Data Integrity and Poisoning Defense:} Ensuring the provenance and integrity of data used to train and fine-tune security LLMs to prevent sophisticated poisoning attacks that could create blind spots or backdoors \cite{korea2023strategy}.
    \item \textbf{Model Evasion and Robustness:} Continuously evaluating and hardening defensive LLMs against adversarial evasion techniques specifically designed to bypass AI-based detection \cite{carlini2021leakage, jia2024global}.
    \item \textbf{Model Confidentiality and Integrity:} Protecting the proprietary architecture and weights of security LLMs from extraction \cite{carlini2021leakage}, and ensuring their operational integrity against unauthorized modifications.
    \item \textbf{Secure Deployment and Monitoring:} Implementing secure deployment practices for LLM-based security tools, including robust access controls, audit trails, and continuous monitoring for anomalous behavior or potential compromise of the AI system itself \cite{singapore2023model}.
\end{itemize}

\section{Future Research Directions}
Building upon the insights and challenges identified in this survey, several critical avenues for future research emerge as essential for advancing the secure and beneficial use of LLMs in cybersecurity. Proactive investigation in these areas will be crucial for staying ahead of evolving threats and harnessing the full defensive potential of these technologies. Key areas warranting dedicated future research include:
\begin{itemize}
    \item \textbf{Developing Novel Robustness Techniques:} Investigating new methods to enhance LLM resilience against sophisticated adversarial attacks, including adaptive defense mechanisms and lifelong learning systems that can evolve with threat landscapes \cite{parisi2019continual}.
    \item \textbf{Scalable and Verifiable Explainability:} Creating XAI techniques for LLMs that are not only interpretable but also verifiable and scalable for complex cybersecurity decision-making processes, ensuring that security analysts can reliably understand and trust LLM outputs \cite{silva2025explainable, mishra2025cysecbench}.
    \item \textbf{Privacy-Preserving LLM Architectures for Security:} Advancing research into novel federated learning configurations, homomorphic encryption applications for LLM inference, and differential privacy guarantees specifically tailored for cybersecurity data and use cases \cite{kairouz2021federated}.
    \item \textbf{Proactive Governance and Ethical Frameworks:} Exploring dynamic and anticipatory governance models that can adapt to the rapid evolution of LLM capabilities and their dual-use implications, including frameworks for continuous ethical impact assessments \cite{european2023ai, brundage2020trustworthy, nist2023rmf, cybersecurity2025trends, china2023interim, japan2022strategy, korea2023strategy, singapore2023model, oecd2023database, unesco2021recommendation}.
    \item \textbf{Cross-Lingual and Multi-Modal Threat Detection:} Enhancing LLM capabilities to detect and analyze threats across diverse languages and data modalities (e.g., code, text, images, network traffic) \cite{geller2024cross} to address the global and multifaceted nature of cyberattacks. This includes leveraging GPU-accelerated feature extraction for real-time vision AI and LLM system efficiency \cite{ahi2025gpu}.
    \item \textbf{Standardized Quantitative Benchmarking:} Establishing more standardized, rigorous, and publicly accessible benchmarks for evaluating the performance, robustness, and fairness of LLMs in diverse cybersecurity tasks, to enable objective comparisons and guide adoption \cite{mishra2025cysecbench, lisha2024zero}. This could also involve AI-powered systems for real-time anomaly detection and data refinement \cite{checkpoint2024sword}.
    \item \textbf{Efficient and Scalable AI Processing:} Investigating architectures for unsupervised, scalable clustering and pattern recognition, potentially leveraging GPU-acceleration, edge, and HPC architectures for challenging high-variability image data relevant to security analytics \cite{ahi2025gpu, carlini2024poisoning}.
\end{itemize}

Finally, industry guidelines and frameworks such as the OWASP Top 10 for LLM Applications \cite{owasp2023top10}, the (ISC)² guidelines on AI and cybersecurity \cite{isc2024navigating}, and the SANS Institute’s white paper on AI-driven security practices \cite{sans2024landscape} further emphasize the urgent need for robust security testing, continuous red-teaming, and practitioner upskilling to counter dual-use threats posed by LLMs.

\section{Conclusion}
Large Language Models (LLMs) are reshaping the cybersecurity landscape—not only by introducing new capabilities for automation, detection, and threat response, but also by amplifying risks through misuse, misalignment, or lack of oversight. As demonstrated across this paper, the same generative power that enables secure code generation, anomaly detection, and zero-day vulnerability identification can also be weaponized to automate malware creation, launch sophisticated phishing campaigns, or bypass traditional defenses.

This dual-use dynamic—the central double-edged sword of LLMs—demands a strategic response that balances innovation with accountability. Defensive use cases such as federated learning, DevSecOps integration, and explainable AI show immense promise, but only when deployed under robust governance structures, ethical auditability, and transparent development practices.

To effectively translate these collaborative imperatives into concrete actions, tailored recommendations for specific stakeholders are essential:
\begin{itemize}
    \item \textbf{For Policymakers and Regulators:} Focus should be on establishing agile and globally harmonized regulatory frameworks that encourage responsible AI innovation while mandating baseline security, transparency, and accountability standards for high-risk LLM applications in cybersecurity. This includes fostering public-private partnerships to share threat intelligence and best practices.
    \item \textbf{For Security Organizations and Practitioners (CISOs, SecOps teams):} Prioritize the development of comprehensive strategies for integrating LLMs into security workflows, including rigorous testing and validation of AI tools, continuous red-teaming exercises against AI-augmented threats, and upskilling security professionals to effectively leverage and manage these technologies.
    \item \textbf{For LLM Developers and AI Researchers:} Emphasize security-by-design principles throughout the LLM lifecycle, from data curation and model training to deployment and monitoring. Invest in research on inherently safer LLM architectures, bias detection and mitigation techniques specific to security contexts, and robust mechanisms for content authenticity and provenance to counter AI-generated disinformation and malware.
\end{itemize}

Ultimately, securing the future of LLMs is not just a technical challenge—it is a societal imperative. Only by embracing both sides of this double-edged sword can we harness the full potential of LLMs to defend the digital frontier while minimizing their risk as instruments of exploitation.

\section*{Acknowledgment}
This paper has been accepted as an invited paper.


\vspace{2cm}

\vspace{2ex}

\begin{IEEEbiography}[{\includegraphics[width=1in,height=1.25in,clip,keepaspectratio]{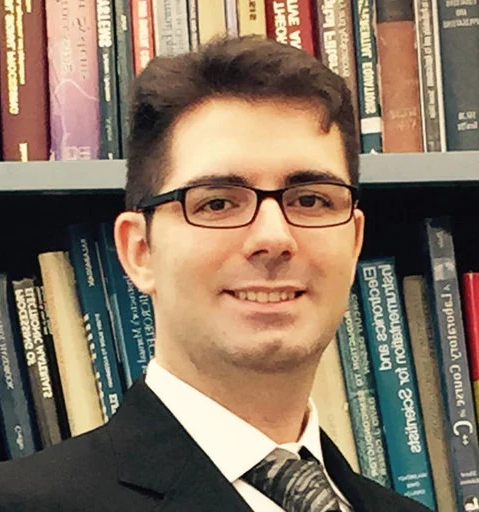}}]{Dr. Kiarash Ahi}
holds M.Sc. and Ph.D. degrees in Electrical and Computer Engineering from Leibniz University Hannover, Germany, and the University of Connecticut, USA, respectively. Dr. Ahi is a 0$\rightarrow$1 product leader, pioneering scientist, distinguished researcher, serial founder, and technology innovator with deep expertise in AI, cybersecurity, Large Language Models (LLMs), Vision-Language Models (VLMs), GPU computing, high-performance computing (HPC) architectures, edge computing, big data analytics, biomedical engineering, digital signal and image processing, natural computation, compressive sensing, optics, and system-level architecture.

Dr. Ahi’s work emphasizes parallel processing, scalable AI models, and intelligent automation, with a strong focus on system design and user experience. His research and industry applications extend across real-time data processing, multimodal intelligence, computer vision, and high-throughput AI platforms.

Dr. Ahi is the founder and CEO of Virelya Intelligence Research Labs, an Adjunct Professor and Graduate Research Co-Advisor at the University of Florida, and has served as an AI Architect for IBM, GlobalFoundries, ASML, and NVIDIA. Since 2019, Dr. Ahi has led the concept-to-market strategy for SEMSuite\textsuperscript{TM}, Siemens' AI-powered analytics platform, as well as Siemens' major digital twins and simulation frameworks, spearheading multinational teams across the globe in orchestrating the full-spectrum scaling of AI-optimized and UX-aware tools including RDF, CDF, CPG, and CMi into scalable, cross-industry solutions optimized for both AI performance and user experience, driving multi-million dollar revenue and receiving multiple performance awards.

Dr. Ahi holds more than 10 patents, has published over 50 peer-reviewed papers, and his work has garnered more than 2,500 citations.

Additionally, Dr. Ahi has developed and monetized more than 10 research-based creative AI-powered applications for iOS, Android, and macOS platforms, achieving more than one million global users.

Dr. Ahi is the recipient of the IEEE AI 1st Place Award and is recognized as a Top Peer-Reviewer by Publons, with over 200 reviews for leading publishers such as Nature, IEEE, Springer, and Elsevier.

As a thought leader in AI ethics and governance, Dr. Ahi advocates for responsible AI deployment, data privacy, and regulatory compliance, shaping the future of digital trust and platform security.

Dr. Ahi has served as a co-advisor to several PhD students and has been an invited IEEE tech speaker on trustworthy AI, LLMs, app safety, platform integrity, automated review systems, advanced imaging systems, and cybersecurity.
\end{IEEEbiography}

\begin{IEEEbiography}[{\includegraphics[width=1in,height=1.25in,clip,keepaspectratio]{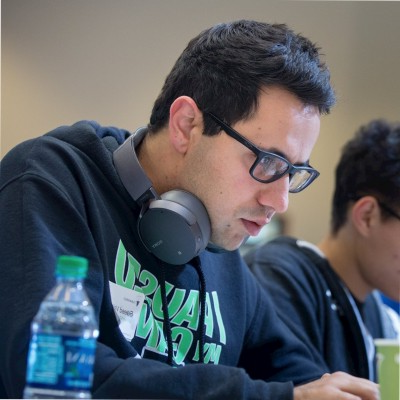}}]{Dr. Saeed Valizadeh}
holds a Ph.D. degree in Computer Science and Engineering from the University of Connecticut. His research focuses on cybersecurity, with an emphasis on modeling attacker-defender interactions using mathematical frameworks. He is currently a lead security researcher at Google.
\end{IEEEbiography}


\begin{thebibliography}{77}

\bibitem{openai2023gpt4} OpenAI, ``GPT-4 Technical Report,'' \emph{arXiv:2303.08774}, 2023.
\bibitem{google2023gemini} Google AI Blog, ``Gemini Overview,'' Google LLC, 2023.
\bibitem{microsoft2023copilot} Microsoft, ``Introducing Security Copilot,'' Microsoft, 2023.
\bibitem{gartner2023risks} Gartner, ``Emerging Security Risks with AI,'' \emph{Gartner Report}, 2023.
\bibitem{narayanan2023dynamic} S. Narayanan et al., ``Dynamic Analysis of Malicious Apps,'' \emph{IEEE Access}, 2023.
\bibitem{pearce2022automated} K. Pearce et al., ``Automated Code Generation Security Risks,'' \emph{IEEE S\&P}, 2022.
\bibitem{chen2023static} J. Chen et al., ``Static Analysis Using LLMs,'' \emph{IEEE Security \& Privacy}, 2023.
\bibitem{carlini2021leakage} N. Carlini et al., ``Risks of LLM Data Leakage,'' \emph{USENIX Security}, 2021.
\bibitem{european2023ai} European Commission, ``Artificial Intelligence Act,'' \emph{EU}, 2023.
\bibitem{brundage2020trustworthy} M. Brundage et al., ``Toward Trustworthy AI Development,'' \emph{arXiv:2004.07213}, 2020.
\bibitem{gitlab2023devsecops} GitLab, ``Auto DevSecOps Powered by AI,'' \emph{GitLab Docs}, 2023.
\bibitem{azure2023sdl} Microsoft Azure, ``Secure Development Lifecycle,'' \emph{Azure Blog}, 2023.
\bibitem{vaithilingam2022cooperative} P. Vaithilingam et al., ``Expecting the Unexpected: Failure Modes of LLMs in Software Engineering,'' \emph{ICSE}, 2022.
\bibitem{microsoft2023copilotcase} Microsoft, ``Security Copilot Case Study,'' Microsoft, 2023.
\bibitem{lisha2024benchmarking} M. Lisha et al., ``Benchmarking LLM for Zero day Vulnerabilities,'' \emph{IEEE CONECCT}, 2024.
\bibitem{kairouz2021federated} P. Kairouz, H. B. McMahan, B. Avent, A. Bellet, M. Bennis, A. Nedić, and U. Ozdaglar, ``Advances and Open Problems in Federated Learning,'' \emph{Found. Trends Mach. Learn.}, vol. 14, no. 1, pp. 1–210, 2021.
\bibitem{bonawitz2019federated} K. Bonawitz, V. Ivanov, B. Kreuter, and S. Marcedone, ``Towards Federated Learning at Scale,'' in \emph{Proc. SysML}, 2019, pp. 1–12.
\bibitem{ribeiro2016why} M. T. Ribeiro et al., ``Why Should I Trust You? Explaining the Predictions of Any Classifier,'' \emph{KDD}, 2016.
\bibitem{samek2021explainable} W. Samek et al., ``Explainable Artificial Intelligence,'' \emph{Springer}, 2021.
\bibitem{jia2024global} X. Jia et al., ``Global Challenge for Safe and Secure LLMs Track 1,'' \emph{arXiv:2411.14502}, 2024.
\bibitem{zhang2020policy} X. Zhang et al., ``Policy Enforcement in App Stores,'' \emph{IEEE TSE}, 2020.
\bibitem{ghaffarinia2022static} R. Ghaffarinia et al., ``Static Android Malware Detection,'' \emph{Elsevier}, 2022.
\bibitem{jiang2019fake} L. Jiang et al., ``Fake Review Detection,'' \emph{ACM CIKM}, 2019.
\bibitem{cybersecurity2025trends} Cybersecurity Ventures, ``Top Cybersecurity Facts, Figures, Predictions, And Statistics For 2025,'' \emph{Cybercrime Magazine}, 2025.
\bibitem{amazon2023automated} Amazon AWS, ``Automated Vulnerability Detection,'' \emph{AWS}, 2023.
\bibitem{ibm2023watsonx} IBM Research, ``Watsonx Security Applications,'' \emph{IBM}, 2023.
\bibitem{palantir2023aip} Palantir, ``AIP for Cybersecurity,'' \emph{Palantir Technologies}, 2023.
\bibitem{ieee2023fairness} IEEE Ethics, ``AI Fairness Guidelines,'' \emph{IEEE}, 2023.
\bibitem{nist2023rmf} NIST, ``AI Risk Management Framework,'' \emph{NIST Special Publication}, 2023.
\bibitem{wef2023standards} World Economic Forum, ``Global AI Security Standards,'' \emph{WEF}, 2023.
\bibitem{google2023safe} Google Security Blog, ``SAFE Framework Implementation,'' \emph{Google}, 2023.
\bibitem{cybersecurity2024almanac} Cybersecurity Ventures, ``Cybersecurity Almanac: 2024 Edition,'' \emph{Cybersecurity Ventures}, Jan. 2, 2024.
\bibitem{google2025gemini} Google Cloud, ``Security, privacy, and compliance for Gemini Code Assist Standard and Enterprise,'' 2025.
\bibitem{microsoft2025copilotfaq} Microsoft, ``Microsoft Security Copilot Frequently Asked Questions,'' 2025.
\bibitem{amazon2025whisperer} Amazon Web Services, ``Use CodeWhisperer to identify issues and use suggestions to improve code security in your IDE,'' 2025.
\bibitem{mulki2025systems} R. Mulki, ``Building Production-Ready LLM Systems: Scaling, Monitoring, and Deployment,'' \emph{Medium}, May 2025.
\bibitem{networks2024fairness} Palo Alto Networks, ``Fairness and Safety of LLMs,'' Jun. 2024.
\bibitem{mohindroo2024privacy} S. Mohindroo, ``Data Privacy and Compliance for Large Language Models (LLMs),'' \emph{Medium}, Sep. 2024.
\bibitem{qualys2025llm} Qualys, ``What is Large Language Model (LLM) Security,'' Apr. 2025.
\bibitem{networks2025explainable} Palo Alto Networks, ``What Is Explainable AI (XAI)?,'' 2025.
\bibitem{zafar2017fairness} M. B. Zafar et al., ``Fairness Constraints: Mechanisms for Fair Classification,'' in \emph{Proc. 20th Int. Conf. Artif. Intell. Statist. (AISTATS)}, Fort Lauderdale, FL, USA, Apr. 2017, pp. 962–970.
\bibitem{barocas2019fairness} S. Barocas, M. Hardt, and A. Narayanan, \emph{Fairness and Machine Learning}, fairmlbook.org, 2019.
\bibitem{ahi2025gpu} K. Ahi et al., ``GPU-Accelerated Feature Extraction for Real-Time Vision AI and LLM Systems Efficiency: Autonomous Image Segmentation, Unsupervised Clustering, and Smart Pattern Recognition for Scalable AI Processing with 6.6$\times$ Faster Performance, 2.5$\times$ Higher Accuracy, and UX-Centric UI Boosting Human-in-the-Loop Productivity,'' \emph{IEEE, ASMC}, Albany, NY, May 2025.
\bibitem{eu2016gdpr} European Union, ``General Data Protection Regulation (GDPR),'' 2016.
\bibitem{ca2018ccpa} State of California Department of Justice, ``California Consumer Privacy Act (CCPA),'' 2018.
\bibitem{ribeiro2020checklist} R. Ribeiro et al., ``Beyond Accuracy: Behavioral Testing of NLP Models with CheckList,'' in \emph{Proc. 58th Annu. Meeting Assoc. Comput. Linguist. (ACL)}, Jul. 2020, pp. 4902–4912.
\bibitem{gunning2017explainable} D. Gunning, ``Explainable Artificial Intelligence (XAI),'' Defense Advanced Research Projects Agency (DARPA), 2017.
\bibitem{ribeiro2016kdd} M. T. Ribeiro, S. Singh, and C. Guestrin, ``Why Should I Trust You?: Explaining the Predictions of Any Classifier,'' in \emph{Proc. 22nd ACM SIGKDD Int. Conf. Knowl. Discov. Data Min.}, San Francisco, CA, USA, Aug. 2016, pp. 1135–1144.
\bibitem{lundberg2017nips} S. Lundberg and S.-I. Lee, ``A Unified Approach to Interpreting Model Predictions,'' in \emph{Proc. 31st Int. Conf. Neural Inf. Process. Syst. (NIPS)}, Long Beach, CA, USA, Dec. 2017, pp. 4765–4774.
\bibitem{desai2025whisper} A. Desai, M. L. Siddiq, and J. C. S. Santos, ``Protecting the Whisper: A Security Assessment of Amazon CodeWhisperer’s Generated Code,'' \emph{ResearchGate}, May 2025.
\bibitem{edpb2025ai} European Data Protection Board, ``AI Privacy Risks and Mitigations – Large Language Models (LLMs),'' Apr. 2025.
\bibitem{silva2025explainable} J. Silva, ``Explainable AI in Cybersecurity: Bridging Transparency and Trust,'' in \emph{Proc. IEEE Conf. Cybersecurity Innovations}, 2025, pp. 78–83.
\bibitem{mishra2025cysecbench} A. Mishra, R. Kumar, and P. Singh, ``CySecBench: A Benchmark Dataset for Evaluating Explainability in Security Contexts,'' \emph{IEEE Trans. Info. Forensics Security}, vol. 20, no. 3, pp. 456–468, 2025.
\bibitem{wang2025cybermentor} F. Wang, L. Zhao, and M. Chen, ``CyberMentor: Enhancing Cybersecurity Learning through Explainable AI,'' in \emph{Proc. IEEE Int. Conf. Emerging Trends in Cyber Training}, 2025, pp. 102–107.
\bibitem{barrett2023ethical} R. Barrett, S. Lee, and T. Harmon, ``Ethical Auditing in AI: The Role of Model Cards and the Cyber Kill Chain,'' \emph{IEEE Trans. Technol. Soc.}, vol. 10, no. 2, pp. 123–132, 2023.
\bibitem{gupta2023framework} P. Gupta, N. Sharma, and K. Desai, ``A Framework for Ethical AI Compliance under the EU AI Act,'' in \emph{Proc. IEEE Workshop on AI Governance}, 2023, pp. 44–49.
\bibitem{kairouz2021federated} P. Kairouz, H. B. McMahan, B. Avent, A. Bellet, M. Bennis, A. Nedić, and U. Ozdaglar, ``Advances and Open Problems in Federated Learning,'' \emph{Found. Trends Mach. Learn.}, vol. 14, no. 1, pp. 1–210, 2021.
\bibitem{bonawitz2019federated} K. Bonawitz, V. Ivanov, B. Kreuter, and S. Marcedone, ``Towards Federated Learning at Scale,'' in \emph{Proc. SysML}, 2019, pp. 1–12.
\bibitem{lisha2024zero} M. Lisha, Y. Zhang, and C. Roberts, ``Benchmarking LLMs for Zero Day Vulnerability Detection,'' in \emph{Proc. IEEE Conf. Emerging Technologies in Security}, 2024, pp. 95–102.
\bibitem{vaithilingam2022cooperative} P. Vaithilingam, A. Deshmukh, and S. Patel, ``Cooperative Human–AI Collaboration for Secure Software Development,'' in \emph{Proc. IEEE Int. Symp. Software Reliability Engineering}, 2022, pp. 215–220.
\bibitem{brundage2020trustworthy} M. Brundage, A. Avin, and S. Clark, ``Accountability in AI: Structured Red Teaming and Model Evaluation Cards,'' \emph{arXiv:2004.07213}, 2020.
\bibitem{parisi2019continual} G. I. Parisi, R. Kemker, J. L. Part, C. Kanan, and S. Wermter, ``Continual lifelong learning with neural networks: A review,'' \emph{Neural Networks}, vol. 113, pp. 54–71, 2019. doi: 10.1016/j.neunet.2019.01.012.
\bibitem{geller2024cross} S. Geller, M. Sivan, R. A. Shkoury, and Y. Elovici, ``Cross-Modal Security: The Impact of Multi-Modal Foundation Models on Industrial Control Systems,'' \emph{arXiv preprint arXiv:2405.11121}, 2024.
\bibitem{kirchenbauer2023watermark} J. Kirchenbauer, J. Geiping, Y. Wen, J. Katz, I. Miers, and T. Goldstein, ``A Watermark for Large Language Models,'' in \emph{Proc. Int. Conf. on Machine Learning (ICML)}, 2023, pp. 11561–11575.
\bibitem{proofpoint2024state} Proofpoint, ``State of the Phish Report,'' Proofpoint Inc., 2024.
\bibitem{checkpoint2024sword} Check Point Research, ``The Double-Edged Sword: How Generative AI is Reshaping Cyber Threats,'' Check Point Research, 2024.
\bibitem{carlini2024poisoning} N. Carlini et al., ``Poisoning Language Models During Instruction Tuning,'' \emph{arXiv preprint arXiv:2403.04557}, 2024.
\bibitem{owasp2023top10} OWASP Foundation, ``OWASP Top 10 for Large Language Model Applications,'' OWASP, 2023.
\bibitem{isc2024navigating} (ISC)², ``Navigating the Intersection of AI and Cybersecurity,'' (ISC)² White Paper, 2024.
\bibitem{sans2024landscape} SANS Institute, ``AI and Cybersecurity: The Evolving Landscape,'' SANS White Paper, 2024.
\bibitem{ahi2025spie} K. Ahi, ``AI-powered end-to-end product lifecycle: UX-centric human-in-the-loop system boosting reviewer productivity by 82\% and accelerating decision-making via real-time anomaly detection and data refinement with GPU-accelerated computer vision, edge computing, and scalable cloud,'' in \emph{Proc. SPIE}, vol. 12782, 2025, Art. no. 1278210. doi: 10.1117/12.1278210.
\bibitem{china2023interim} Cyberspace Administration of China, ``Interim Measures for the Management of Generative Artificial Intelligence Services,'' Beijing, China, 2023.
\bibitem{japan2022strategy} Cabinet Office, Government of Japan, ``AI Strategy 2022,'' Tokyo, Japan, 2022.
\bibitem{korea2023strategy} Ministry of Science and ICT, Republic of Korea, ``National Strategy for Artificial Intelligence,'' Seoul, Korea, 2023.
\bibitem{singapore2023model} Infocomm Media Development Authority (IMDA) Singapore, ``Model AI Governance Framework,'' 2023.
\bibitem{oecd2023database} OECD.AI Policy Observatory, ``Database of National AI Strategies and Policies,'' OECD, 2023.
\bibitem{unesco2021recommendation} UNESCO, ``Recommendation on the Ethics of Artificial Intelligence,'' Paris, France, 2021.

\end{thebibliography}
\end{document}